\documentclass[11pt,twoside]{article}

\usepackage{asp2006}
\usepackage{epsf}
\usepackage{psfig}
\usepackage{lscape}

\begin{document}
\title{Microlensing Searches for Planets: Results and Future Prospects}   
\author{B. Scott Gaudi}   
\affil{Department of Astronomy, The Ohio State University, 140 W. 18th Ave., Columbus, OH 43210}    %
\begin{abstract} 
Microlensing is 
potentially sensitive to multiple-planet systems containing analogs of
all the solar system planets except Mercury, as well as to free
floating planets. I review the landscape of microlensing planet searches, 
beginning with an outline of the method itself,
and continuing with an overview of the results that have been obtained to
date. Four planets have been detected with microlensing. I
discuss what these detections have taught us about the frequency of
terrestrial and giant planets with separations beyond the ``snow
line.''  I then discuss the near and long-term prospects
for microlensing planet searches, and in particular speculate on the
expected returns of next-generation microlensing experiments both from
the ground and from space. When combined with the results from other
complementary surveys, next generation microlensing surveys can yield
an accurate and complete census of the frequency and properties of
essentially all planets with masses greater than that of Mars. 
\end{abstract}

\section{Introduction}   

\centerline{{\it``I don't understand. You're looking for planets you can't}}
\centerline{{\it see around stars you can't see.''} -- Debra Fischer, c.\ 2000}
\centerline{}
\centerline{{\it``Microlensing is a cult.''} -- Dave Koerner, c.\ 2000}
\centerline{}

At the time when these words were uttered, microlensing planet searches
had not yet made any definitive detections, despite having
been in production in earnest for over five years, and despite the fact that
numerous talks had been given vociferously extolling the virtues of
the method.  Furthermore, it was a well-known `fact' (in part
perpetuated by those giving said talks [including myself]) that the primary virtue of the
microlensing method was its ability to provide statistics on
large-separation and low-mass planets.  Individual detections would be
of little interest, it was thought, because there would
be no opportunity for follow-up and the  properties of the host stars would
remain unknown, due primarily to their very large distances from Earth. 

It is now seven years later, and the landscape and perception of microlensing planet
searches has changed dramatically.  Four 
detections have been published, and an additional $\sim$half a
dozen planets have been detected and will be announced in the
not-to-distant future.  The four published detections have already
provided new and important information about the frequency of ``Super
Earths'' and Jovian planets with separations beyond the ``snow line.''
Furthermore, it has become clear that the original assessments of the
(im)possibility of follow-up and characterization of the planet host stars
were overly pessimistic.  Indeed,
for three of the four detected planets, reasonably precise ($\sim
20\%$) estimates of the mass of the host star have been made or will
soon be possible.

\section{How Microlensing Finds Planets} 

A microlensing event occurs when a
foreground ``lens'' happen to pass very close to our line of sight to a 
more distant background ``source.'' For our purposes, the lens
is a main-sequence star or stellar remnant in the foreground Galactic disk
or bulge, whereas the source is a main-sequence
star or giant typically in the bulge.  If the lens is exactly aligned with the
source, it images the source into an ``Einstein ring'' with a radius 
$\theta_{\rm E}=\sqrt{\kappa M \pi_{\rm rel}}$,
where $M$ is the mass of the lens, $\pi_{\rm rel}={\rm AU}/(d_l^{-1}-d_s^{-1})$ is the lens-source relative
parallax, $d_l$ and $d_s$ are the distances to the source and lens, and 
$\kappa=4G/c^2{\rm AU}$.  In the case of imperfect alignment,
the lens creates two images of the source
which have splittings near the time of closest alignment of $\sim 2\theta_{\rm E}$.
For typical lens masses ($0.1-1~M_\odot$) and lens and source distances ($1-10~{\rm kpc}$),
$\theta_{\rm E}\sim 500~\mu{\rm as}$ and so the images are not resolved.  However, the images
are also magnified, by an amount $A$ that depends only on $u\equiv \theta/\theta_{\rm E}$,
the angular separation of the lens
and source $\theta$ in units of $\theta_{\rm E}$.
For $u \ll 1$, the magnification takes on the simple form $A \simeq 1/u$.

The transverse motion of the lens, source, and observer results in a
time-variable magnification and gives rise to a smooth, symmetric
microlensing event with a characteristic form.  
The magnification is $A>1.34$ for $u\le 1$, and so the
magnifications are substantial and easily detectable.  Typically,
observations of a microlensing event caused by an isolated lens can
fit by a simple four parameter model. Three specify the magnification
as a function of time: the minimum angular separation $u_0$ of the
lens and source in units of $\theta_{\rm E}$ (the impact parameter,
which also specifies the maximum magnification), the time of maximum
magnification, and finally the Einstein timescale, $t_{\rm E}\equiv
\theta_{\rm E}/\mu$, where $\mu$ is the relative lens-source proper
motion.  The typical timescales for events toward the Galactic bulge
are of order a month.  Only $t_{\rm E}$ contains any
information about the physical properties of the lens, and then only in a
degenerate combination of the lens mass, distance, and transverse
velocity.  However, as mentioned below, it has proven to be the case
that it is often possible to obtain additional information which
partially or totally breaks this degeneracy.  The remaining two
parameters required to fit a single-lens event are the flux of the
source, and the flux of any unresolved light that is not being lensed.
The latter can include light from a companion to the source, light from
unrelated nearby stars, light from a companion to the lens, and (most
interestingly) light from the lens itself.  In those cases where it is
possible to isolate the light from the lens itself, this measurement
can be used to constrain the lens mass \citep{bennett07}.

If the lens star happens to have a planetary companion, the companion
can create a short-timescale perturbation on the primary microlensing
event \citep{mao91,gould92}.  There are two conceptually different channels 
by which planets can be detected with microlensing.

In the main channel, the planet happens to be located near the
path of one of the two images created by primary lens.  As the image
sweeps by the position of the planet, the planet will further perturb
the light from this image and yield a short-timescale deviation
\citep{gould92}. The
duration of this deviation is $t_p \sim q^{1/2} t_{\rm E}$,
where $q=m_p/M$ is the mass ratio and $m_p$ is the planet mass. Thus
the duration of the perturbation ranges from a few hours for an
Earth-mass planet to a day for Jovian-mass planets.  The location of
the perturbation relative to the peak of the primary event depends on
the angle of the projected star-planet axis, as well as the
instantaneous angular separation between the planet and host star in
units of $\theta_{\rm E}$.  Since the orientation of the source
trajectory relative to the planet position is random, the time of this
perturbation is not predictable and the probability is $\sim A(t_p)
\theta_p/\theta_{\rm E}$, where $A(t_p)$ is the magnification at the
time $t_p$ of the perturbation, and $\theta_p \equiv q^{1/2}
\theta_{\rm E}$ is the Einstein ring radius of the planet.  The
detection probabilities (given the existence of a planet) range from
tens of percent for Jovian planets to a few percent for Earth-mass
planets \citep{gould92,bennett96}.  Since the planet must be located near one of the two primary
images in order to yield a detectable deviation, and these images are
always located near the Einstein ring radius when the source is
significantly magnified, the sensitivity of the microlensing method
peaks for planet-star separations of $\sim \theta_{\rm E}d_l$.  Detecting
planets via the main channel requires substantial commitment
of resources because the unpredictable nature of the perturbation requires 
dense, continuous sampling, and furthermore the detection probability per event is relatively low
so many events must be monitored.  

The other channel by which microlensing can detect planets is in high
magnification events \citep{griest98}.  In addition to perturbing images that happen to
pass nearby, planets will also distort the perfect circular symmetry
of the Einstein ring.  Near the peak of
high-magnification events, as the lens passes very close to the
observer-source line of sight (i.e.\ when $u\ll 1$), the two primary
images are highly elongated and sweep along the Einstein ring, thus
probing this distortion.  For very high-magnification events ($A\ga
100$), these images probe nearly the entire Einstein ring radius and
so are sensitive to all planets with separations near $\theta_{\rm
E}d_l$, regardless of their orientation with respect to the source
trajectory.  Thus high-magnification events can have nearly 100\%
sensitivity to planets near the Einstein ring radius, and 
are very sensitive to low-mass planets \citep{griest98}.  However, these
events are rare: a fraction $\sim 1/A$ of events have maximum
magnification $\ga A$.  However, these events can often be predicted
several hours to several days ahead of peak, and furthermore the
times of high sensitivity to planets are within a
full-width half-maximum of the event peak,
or roughly a day for typical high-magnification events \citep{rattenbury02}.  
Thus scarce observing resources can be concentrated on these few events and
only during the times of maximum sensitivity.  Because the source
stars are highly magnified, it is also possible to use more common,
smaller-aperture telescopes.

\subsection{Peculiarities and Features of the Microlensing Method}

The unique way in which microlensing finds planets leads to some
useful features, as well as some (mostly surmountable) drawbacks.
Most of the features of the microlensing method can be understood
simply as a result of the fact that planet detection relies on the
direct perturbation of images by the gravitational field of the
planet, rather than on light from the planet, or the indirect effect
of the planet on the parent star.\\
\noindent$\bullet$ {\bf Peak Sensitivity Beyond the Snow Line.}
The peak sensitivity of microlensing is for planet-star separations of 
$\sim \theta_{\rm E}d_l$, which corresponds to equilibrium temperatures of
\begin{equation}
T_{\rm eq} = 278~{\rm K} \left(\frac{L}{L_\odot}\right)^{1/4} \left(\frac{\theta_{\rm E}d_l}{{\rm AU}}\right)^{-1/2}
\sim 70~{\rm K} \left(\frac{M}{0.5M_\odot}\right),
\label{eqn:Teq}
\end{equation}
where I have assumed $L/L_\odot = (M/M_\odot)^5$, $d_l=4~{\rm kpc}$, and $d_s=8~{\rm kpc}$.
Microlensing is most sensitive to planets in the regions beyond the `snow line,' the 
point in the protoplanetary disk
exterior to which the temperature is less than the condensation temperature of water
\citep{lecar06,kennedy07}. Giant planets are thought to form
in the region immediately beyond the snow line, where the 
surface density of solids is highest.

Microlensing is not sensitive to planets
with separations much smaller than the Einstein ring radius, as these can only 
perturbed highly demagnified images.  Thus microlensing is much less
sensitive to planets in the habitable zones of their parent stars \citep{park06}.\\
\noindent$\bullet$ {\bf Sensitivity to Low-mass Planets}
The amplitudes of the perturbations caused by planets are typically large,
$\ga 10\%$,  Furthermore,
although the durations of the perturbations get shorter with planet
mass (as $\sqrt{m_p}$)
and the probability of detection decreases (also as $\sqrt{m_p}$), 
the amplitude of the perturbations are independent of the planet mass.
This holds until the `zone of influence' of the planet, which has a
size $\sim \theta_p$, is smaller than the angular size of the source
$\theta_*$.  When this happens, the perturbation is `smoothed'
over the source size.  For typical parameters, $\theta_p\sim \mu{\rm
as}(m_p/M_\oplus)^{1/2}$, and for a turn-off star in the bulge
$\theta_*\sim \mu{\rm as}$, so this `finite source' suppression begins
to become important for planets with the mass of the Earth, but does
not completely suppress the perturbations until masses below that of
Mars for main-sequence sources \citep{bennett96}.\\
\noindent$\bullet$ {\bf Sensitivity to Long-Period and Free-Floating
Planets}. Since microlensing can `instantaneously' detect planets without waiting
for a full orbital period, it is sensitive to planets with very long
periods.  Although the probability of detecting a planet decreases
for planets with separations larger than the Einstein ring radius
because the magnifications of the images decline, it does not drop to
zero.  Indeed since microlensing is directly sensitive to the planet
mass, planets can be detected even without a primary microlensing event.  
Even free-floating planets that
are not bound to any host star are detectable in this way \citep{han05}.
Microlensing is the only method that can detect old, free-floating
planets.  A significant population of such planets is a
generic prediction of most planet formation models, particular those
that invoke strong dynamical interactions to explain the observed
eccentricity distribution of planets \citep{goldreich04,juric07,ford07}.\\
\noindent $\bullet$ {\bf Sensitivity to Planets Throughout the
Galaxy}.  Because microlensing does not rely on light from the planet
or host star, planets can be detecting orbiting stars with distances
of several kiloparsecs.  The host stars probed by microlensing are
simply representative of the population of massive objects along the
line of sight to the bulge sources, weighted by the lensing
probability.  The lensing probability peaks for lens distances about
halfway to the sources in the Galactic bulge, but remains substantial
for lens distances in the range $d_l\sim 1-8~{\rm kpc}$. Specialized
surveys may be sensitive to planets with $d_l \la 1~{\rm kpc}$ \citep{gaudi07},
as well as planets in M31 \citep{covone00}.\\
\noindent $\bullet$ {\bf Sensitivity to Planets Orbiting a Wide Range
of Host Stars}.  The sensitivity of microlensing is weakly
dependent on the host star mass, and has essentially no dependence on
the host star luminosity.  Thus microlensing is about equally
sensitive to planets orbiting stars all along the main sequence, from
brown dwarfs to the main-sequence turn-off, as well as 
planets orbiting white dwarfs, neutron stars, and black
holes \citep{gould00}.\\
\noindent $\bullet$ {\bf Sensitivity to Multiple-Planet Systems}.  For
the main detection channel, multiple planets in the same system
can be detected only if both planets happen to have projected
positions sufficiently close to the paths of the two images created by
the primary lens.  The probability of this is simply the product of the
individual probabilities, or ${\cal{O}}(1\%)$ \citep{han02}.  In
high-magnification events, however,
individual planets are detected with near-unity
probability regardless of the orientation of the planet with respect to
the source trajectory. This immediately implies all planets
sufficiently close to the Einstein ring radius will be revealed in
such events \citep{gaudi98}.  This, along with the fact that high-magnification events
are potentially sensitive to very low-mass planets, makes such events excellent probes
of planetary systems. 

\subsection{What Can We Learn About the Planets and Host Stars?}

It has often been stated that the ability of microlensing to provide detailed information
about individual systems is very limited. This perception comes from the fact that (1) the
host stars are typically 
distant and faint, making follow-up work is difficult, (2) in the
overwhelming majority of microlensing events, the only parameter that
can be constrained which contains any information about the primary lens is
the event timescale $t_{\rm E}$, which is a degenerate combination of mass,
distance, and transverse velocity of the lens, (3) microlensing detections
routinely provide only the mass ratio of the planet and host star \citep{gaudi97}, and so the mass
of the planet is typically not known without a constraint on the primary mass, and 
furthermore (4) the only constraint on the planet orbit is $b_\perp$, 
the instantaneous angular separation between
the planet and host star at the time of the event in
units of $\theta_{\rm E}$ \citep{gaudi97}.   Since $\theta_{\rm E}$, the inclination, phase, and ellipticity
of the orbit are all unknown, $b_\perp$ provides very
little information about the orbit.  In the cases when only 
$t_{\rm E}$, $q$, and $b_\perp$ can be measured, constraints on the 
mass and distance to the lens, and mass and semimajor axis of the planet,
must rely on a Bayesian analysis which incorporates priors on the distribution
of microlens masses, distances and velocities (e.g., \citealt{dominik06,dong06}).

Experience has shown that, in reality, much more information
can typically be gleaned from a combination of a
detailed analysis of the light curve and follow-up, high-resolution
imaging.  Several additional pieces of information are potentially
available.\\
\noindent $\bullet${\bf Finite Source Effects}: For the majority of
planets detected via microlensing, the `smoothing' effects of the
finite source size are detectable during sharp features in the
light curve caused by the planet.  The magnitude of this effect
is set by $\rho_*\equiv \theta_*/\theta_{\rm E}$, and since $\theta_*$
can be inferred from the color and magnitude of the source, this
allows a measurement of $\theta_{\rm E}$ \citep{gould94}.\\
\noindent $\bullet${\bf Microlens Parallax}: In at least three of the
planetary microlensing events, it has also been possible to measure
the deviations in the microlensing light curve caused by the fact that
the Earth is accelerating \citep{gould92a}.  These deviations are generally only
significant for events with timescales that are a significant fraction
of a year, and so long as compared to the median timescale of $\sim
20~{\rm days}$.  However, due to selection
effects, the majority of planetary events have been long
timescale. This `microlens parallax' allows one to constrain $\tilde
r_{\rm E}$, the Einstein ring radius
projected onto the observer plane.\\
\noindent $\bullet${\bf Light from the Lens}: Although the lenses are
expected to be distant and low-mass, and so faint, it is nevertheless
possible to detect some main-sequence lenses during
and/or after the event and measure their flux to reasonable precision
(often in several different filters).  This allows for a photometric
estimate of the lens mass and distance \citep{bennett07}.\\
\noindent $\bullet${\bf Proper Motion of the Lens}: For microlensing
events toward the Galactic bulge, the relative lens-source proper
motions are $\mu_{\rm rel} \sim 5-10~{\rm mas/yr}$.  Thus, after $\sim 5-10$ years,
the lens and source will be displaced by $\sim 0.05$ arcseconds.
For luminous lenses, and using space telescope or adaptive optics imaging, it is possible
measure the relative lens-source proper motion, either by measuring
the elongation of the PSF or by measuring the difference in the
centroid in several filters if the lens and source have significantly
different colors \citep{bennett07}.  The proper motion can be combined 
with the timescale to give the Einstein ring radius, $\theta_{\rm E}=\mu_{\rm rel} t_{\rm E}$.\\
\noindent $\bullet${\bf Orbital Motion of the Planet}: In at least two
cases, the orbital motion of the planet during the microlensing event
has been detected.  The effects of orbital motion generally allow the
measurement of the two components of the projected velocity of the
planet relative to the primary star. If an external measurement of the
mass of the lens is available, and under the assumption of a circular
orbit, these two components of the projected velocity completely
specify the full orbit of the planet (including inclination), up to a
two-fold degeneracy \citep{dong08}. In some unusual cases, higher-order effects of
orbital motion can be used to break this degeneracy and even constrain
the ellipticity of the orbit.

In many cases, several of these pieces of information can be measured
in the same event, often providing complete or even redundant measurements
of the mass, distance, and transverse velocity of the event.  For example,
a measurement of $\theta_{\rm E}$ from finite source effects, when combined
with a measurement of $r_{\rm E}$ from microlens parallax, yields
the lens mass $M=(c^2/4G)\tilde r_{\rm E} \theta_{\rm E}$, distance
$d_{l}^{-1} = \theta_{\rm E}/\tilde r_{\rm E} + d_s^{-1}$,
and transverse velocity \citep{gould00b}.

\begin{figure}[ht]
\plottwo{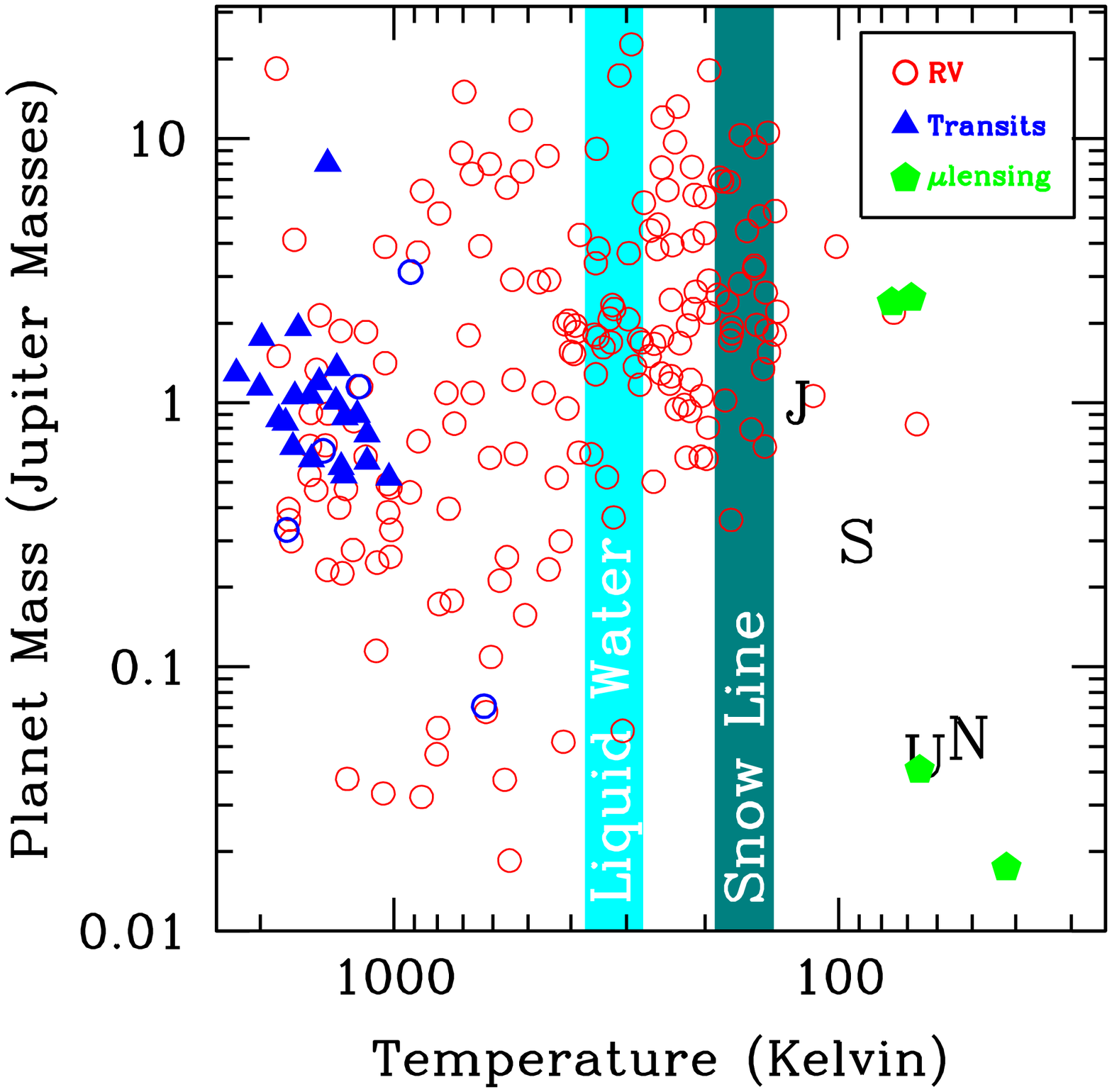}{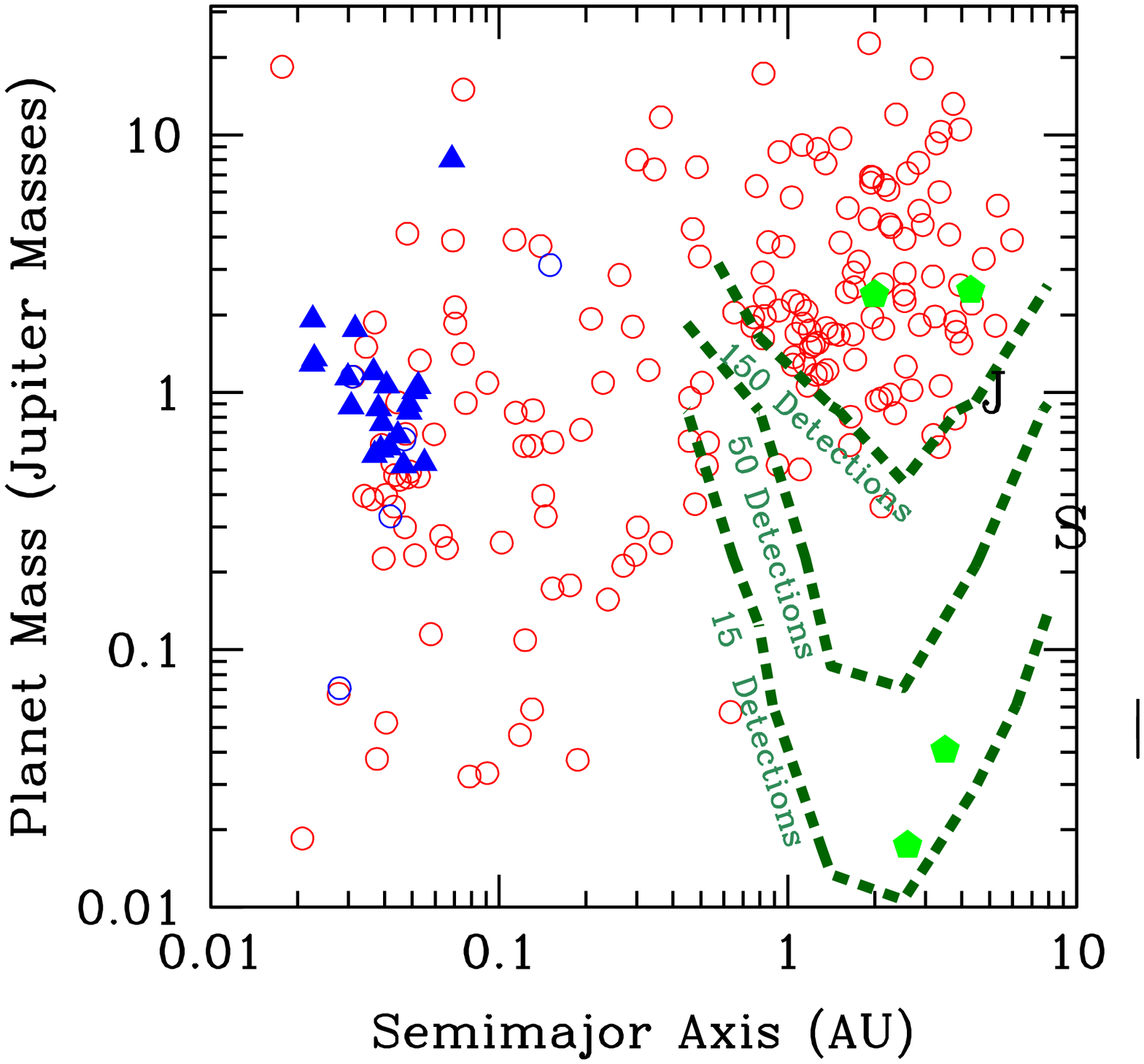}
\caption{
{\bf Left} Extrasolar planets detected via 
transits (triangles), RV (circles), and microlensing (hexagons), as a function
of their mass and equilibrium temperature. Blue circles are planets
found via RV that were subsequently found to be transiting.
Also shown are the approximate locations of the habitable zone
(e.g., \citealt{kasting93}) and the snow line (e.g., \citealt{kennedy07,lecar06}).
{\bf Right} Same as
the right panel, but versus semimajor axis.  The contours show
the anticipated number of detections per year from a next-generation microlensing survey,
assuming all stars have planet with the given mass or semimajor axis.  
}
\end{figure}

\section{How Microlensing Planet Searches Work (or Don't) in Practice}

The microlensing event rate toward the Galactic bulge is
${\cal O}(10^{-6})$ events per star per year.  In a typical field toward
the Galactic bulge, the surface density of stars is $\sim 10^{7}$
stars per deg$^2$ to $I\sim 20$.  Thus to detect $\sim 100$
events per year, $\sim 10$ deg$^2$ of the Bulge must be
monitored.  Current large-format CCD cameras typically have fields of
view of $\sim 0.25$ deg$^2$, and thus $\sim 40$ pointings are
required and so fields can only be monitored once or twice per
night.  While this cadence is sufficient to detect
the primary microlensing events, it is
insufficient to detect and characterize planetary perturbations, which
last a few days or less.

Thus microlensing planet searches currently operate using a two-stage
process.  The Optical Gravitational Lens Experiment (OGLE, \citealt{udalski02}) and the
Microlensing Observations in Astrophysics (MOA, \citealt{abe04}) collaborations monitor
several tens of square degrees of the Galactic bulge, reducing their
data real-time in order to alert microlensing event in progress.  A
subset of these alerted events are then monitored by two follow-up
collaborations, a joint venture of the Probing Lensing Anomalies
NETwork (PLANET, \citealt{albrow98}) and RoboNet \citep{burgdorf07}
collaborations, and the Microlensing Follow Up Network ($\mu$FUN,
\citealt{yoo04}).  Since only individual events are monitored, these
teams can achieve the sampling and photometric accuracy necessary to
detect planetary perturbations.  In fact, the line between the `alert' and `follow-up'
collaborations is somewhat blurry, both because the MOA and OGLE
collaborations monitor some fields with sufficient cadence to detect a
subset of the longest planetary perturbations, and because there is a
high level of communication between the collaborations, such that the
observing strategies are often altered real
time based on available information about ongoing events.

The PLANET/RoboNET collaboration has substantial access to 0.6-1.5m
telescopes located in Chile, South Africa, Perth, and Tasmania.  With
these resources, they are able to monitor dozens of events
per season, and so are able to search for planets via the main channel.  
This tactic led to the detection of the
first cool rocky/icy exoplanet OGLE-2005-BLG-390Lb \citep{beaulieu06}.

The $\mu$FUN collaboration takes a different approach, motivated by
its more shoe-string, slipshod nature.  They use a single 1m telescope
in Chile to monitor promising alerted events in order try to identify
high-magnification events substantially before peak.  Since
high-magnification events are relatively rare ($\sim 5$ per year),
for the majority of the time, nothing happens.  However, when a
likely high-magnification event is imminent, they notify the other telescopes
in the collaboration, and then instruct them to `go all out' during the high-magnification
peak.  Thus $\mu$FUN adopts what can best be described as a  ``wait... wait... wait... PANIC'' 
approach to searching for planets.  Since $\mu$FUN focuses on high-magnification
events, which often reach peak magnitudes of $I \la 15$, their targets are
usually within reach of the relatively small-aperture (0.3-0.4m) telescopes commonly
owned by enthusiastic amateur astronomers.  Indeed, over half of the members
of the $\mu$FUN collaboration are amateurs. These amateurs often contribute
crucial data: in the majority of the $\mu$FUN planet detections, data from amateur telescopes
were essential for the proper interpretation of the events.  In the words
of one amateur member of $\mu$FUN:\\

\centerline{{\it``It just shows that you can be a mother, you can work full-time, and you can}} 
\centerline{{\it still go out there and find planets.''} -Jenny McCormick, Farm Cove Observatory}
\centerline{}

The real-time identification of high-magnification events is a difficult business, and
is often prone to human error, as well as human inspiration.  For example, in the case of 
microlensing event OGLE-2004-BLG-343, an internal alert
was ignored and as a result the peak of this magnification $\sim 3000$ event was missed.  Subsequent
analysis demonstrated that had the peak been intensely monitored, 
the event would have provided significant sensitivity to Earth-mass planets \citep{dong06}.  On the other hand, in the case
of the magnification $\sim 800$ event OGLE-2005-BLG-169, the specific instructions of a senior scientist 
to gather a few data points were explicitly ignored by an insubordinate graduate student,
and instead said graduate student obtained over 1000 points over the peak, resulting in 
the detection of a Neptune-mass planet \citep{gould06}.

\section{What We've Found: Results to Date} 

The microlensing method was first proposed in 1991, and microlensing
planet searches began in earnest in 1995.  From 1995-2001, 
only $\sim 50-100$ events were alerted per
year. As a result, there were few high magnification events each year,
and only a handful of events ongoing at any time.  The follow-up
collaborations were therefore not able to `pick and choose' the best
events for follow-up.  Indeed, no convincing planet detections were
made during this period, although interesting upper limits were placed
on the frequency of Jovian planets based on detailed analyses of the
sensitivities of the various searches \citep{gaudi02,snodgrass04}.  Perhaps the most important
result of this period, however, was the development of both the
theory and practice of the microlensing method, which resulted in its
transformation from a theoretical abstraction to a viable, practical
method of searching for planets.

In 2001, the OGLE collaboration upgraded to a new camera with a 16
times larger field of view and so were able to monitor a
larger area of the bulge with a higher
cadence.  As a result, in 2002 OGLE began alerting nearly an order
of magnitude more events per year than previous to the upgrade.  These
improvements in the alert rate and cadence, combined with improved
cooperation and coordination between the survey and follow-up
collaborations, led to the first discovery of an extrasolar planet
with microlensing in 2003.  MOA upgraded to a 1.8m telescope and 2 deg$^2$
camera in 2004, and together OGLE and MOA collaborations
now alert $\sim 850$ events per year.

To date, there are four published planet detections with microlensing.
The masses, separations, and equilibrium temperatures of these
planets are shown in Figure 1.  The first two planets found by
microlensing, OGLE-2003-BLG-235/MOA-2003-BLG-53Lb \citep{bond04}, and
OGLE-2005-BLG-071Lb \citep{udalski05}, are Jovian-mass objects with separations of $\sim
2-4~{\rm AU}$. While the masses and separations of these planets
are similar to many of the planets discovered via radial velocity
surveys, their host stars are generally less massive and so the
planets have substantially lower equilibrium temperatures of $\sim 70$~K,
similar to Saturn and Uranus.

The third and fourth planets discovered by microlensing are
significantly lower mass, and indeed inhabit a region of
parameter space that was previously unexplored by any
method.  OGLE-2005-BLG-390Lb is a very low-mass planet with a
planet/star mass ratio of only $\sim 8 \times 10^{-5}$ \citep{beaulieu06}.  A Bayesian
analysis combined with a measurement of $\theta_{\rm E}$ from finite
source effects indicates that the planet likely orbits a low-mass M
dwarf with $M=0.22_{-0.11}^{+0.21}M_\odot$, and thus has a mass of
only $5.5_{-2.7}^{+5.5}M_\oplus$.  Its separation is
$2.6_{-0.6}^{+1.5}~{\rm AU}$, and so has an extremely cool equilibrium
temperature of $\sim 50~{\rm K}$.  OGLE-2005-BLG-169Lb is another
low-mass planet with a mass ratio of $8\times 10^{-5}$ \citep{gould06}, essentially
identical to that of OGLE-2005-BLG-390Lb.  A Bayesian
analysis indicates a primary mass of $0.52_{-0.22}^{+0.19}M_\odot$,
and so a planet of mass $\sim 14_{-6}^{+5} M_\oplus$, a separation
of $3.3_{-0.9}^{+1.9}~{\rm AU}$, and so an equilibrium temperature of
$\sim 70~{\rm K}$.  In terms of its mass and equilibrium temperature,
OGLE-2005-BLG-169Lb is very similar to Uranus.  OGLE-2005-BLG-169Lb
was discovered in a high-magnification $A\sim 800$ event; as argued
above, such events have significant sensitivity to multiple
planets.  There is no
indication of any additional planetary perturbations in this event,
which excludes Jupiter-mass planets with separations between 0.5-15~AU,
and Saturn-mass planets with separations between 0.8-9.5~AU.  Thus
it appears that this planetary system
is dominated by the Neptune-mass companion.  

The microlensing detection sensitivity declines with planet mass as
$\sqrt{m_P}$, and thus the presence of two low-mass planets in a sample of
four detections argues that the frequency of cool
``Neptunes'' ($5-15~M_\oplus$) is substantially higher than that of
cool Jovian-class planets.  A quantitative analysis that accounts for
the detection sensitivities and Poisson statistics shows that, at 90\%
confidence, $38_{-22}^{+31}\%$ of stars host cool
Neptunes with separations in the range $1.6-4.3~{\rm AU}$ \citep{gould06}.  Thus,
these planets are common, which is ostensibly a confirmation of the
core accretion model of planet formation, which predicts that there
should exist many more `failed Jupiters' than bona-fide Jovian-mass
planets at such separations, particularly around low-mass primaries.

In all of the planet detections, it has been possible to obtain
additional information to improve the constraints on the properties of
the primaries and planets.  In all four cases, finite source effects
have been detectable and so it has been possible to measure
$\theta_{\rm E}$.  For OGLE-2003-BLG-235/MOA-2003-BLG-53Lb, follow-up
imaging with the {\it Hubble Space Telescope (HST)} yielded a
detection of light from the lens, which constrains the mass of the
primary to $\sim 15\%$, $M=0.63_{-0.09}^{+0.07}M_\odot$ \citep{bennett06}. In the case
of OGLE-2005-BLG-071Lb, {\it HST} photometry, when combined with
information on finite source effects and
microlens parallax from the light curve, leads to the conclusion that the primary lens is a
mid-M dwarf with a mass of $M=0.35 \pm 0.05 M_\odot$ and a distance of
$d_l=3.1 \pm 0.4~{\rm kpc}$ \citep{dong08}.  Furthermore, the velocity of the primary
is constrained to be $v_{\rm LSR}=110\pm 25~{\rm km~s^{-1}}$ and there
evidence at the $\sim 2\sigma$ level that the host has a subsolar
metallicity.  Thus, OGLE-2005-BLG-071Lb
may be a massive Jovian planet orbiting a metal-poor, thick-disk, mid
M-dwarf.  The existence of such a planet may pose a challenge for core-accretion models
of planet formation (e.g., \citealt{laughlin04,ida04,ida05}).  
Although this result should be taken as
tentative, and must be confirmed with additional measurements, it
nevertheless demonstrates that it is possible to obtain fairly
detailed information on individual planetary systems detected via
microlensing.

\section{Where We Are Going: the Short and Long-Term Future}

With the recent MOA upgrade, the rate of planet
detections has increased substantially.  From 2003-2006, six planets
were detected (four have been published).  From the 2007
bulge season alone, there are four fairly secure planetary events.
This rate can be expected to increase modestly as analysis techniques
improve, and so the next several years should bring of order a dozen
planet detections.

The new MOA setup points the way toward the next generation of
microlensing planet searches, which will operate in a very different
mode than the current alert/follow-up model.  With a
sufficiently large field-of-view (FOV) of 2-4 deg$^2$, it becomes
possible to monitor tens of
millions of stars every $10-20$ minutes, and so discover thousands of
microlensing events per year. Furthermore, these would be
simultaneously monitored to search for planetary perturbations.  In
order to obtain round-the-clock coverage and so catch all of the
perturbations, several such telescopes would be needed, located on 3-4
continents roughly evenly spread in longitude. 
Detailed simulations of such a next-generation microlensing survey
have been performed by several groups.  These simulations include
models for the Galactic population of lenses and sources that match all
constraints, and account for real-world effects such as weather,
variable seeing, moon and sky background, and crowded fields.  They
reach similar conclusions.  Such a survey would increase the planet
detection rate at fixed mass by at least an order of magnitude over
current surveys.  Figure 2 shows the predictions of these simulations
for the detection rate of planets of various masses and
separations.  In particular, if Earth-mass planets with semimajor axes
of several AU are common around main-sequence stars, a next generation
microlensing survey should detect several such planets per year.
This survey would also be sensitive to free-floating planets, and would detect
them at a rate of hundreds per year if every star has ejected Jupiter-mass
planet \citep{gould07}.

\begin{figure}[ht]
\plottwo{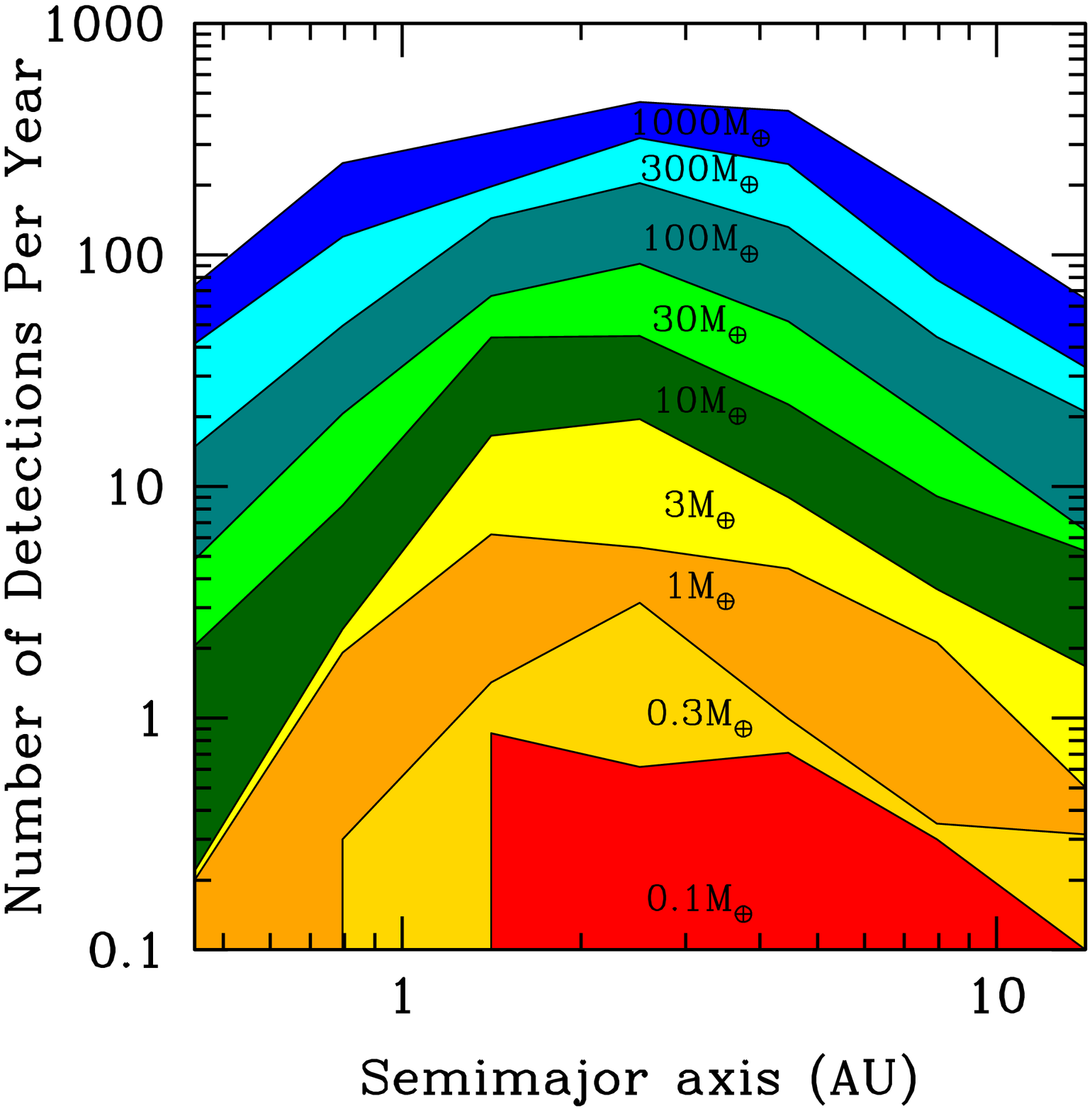}{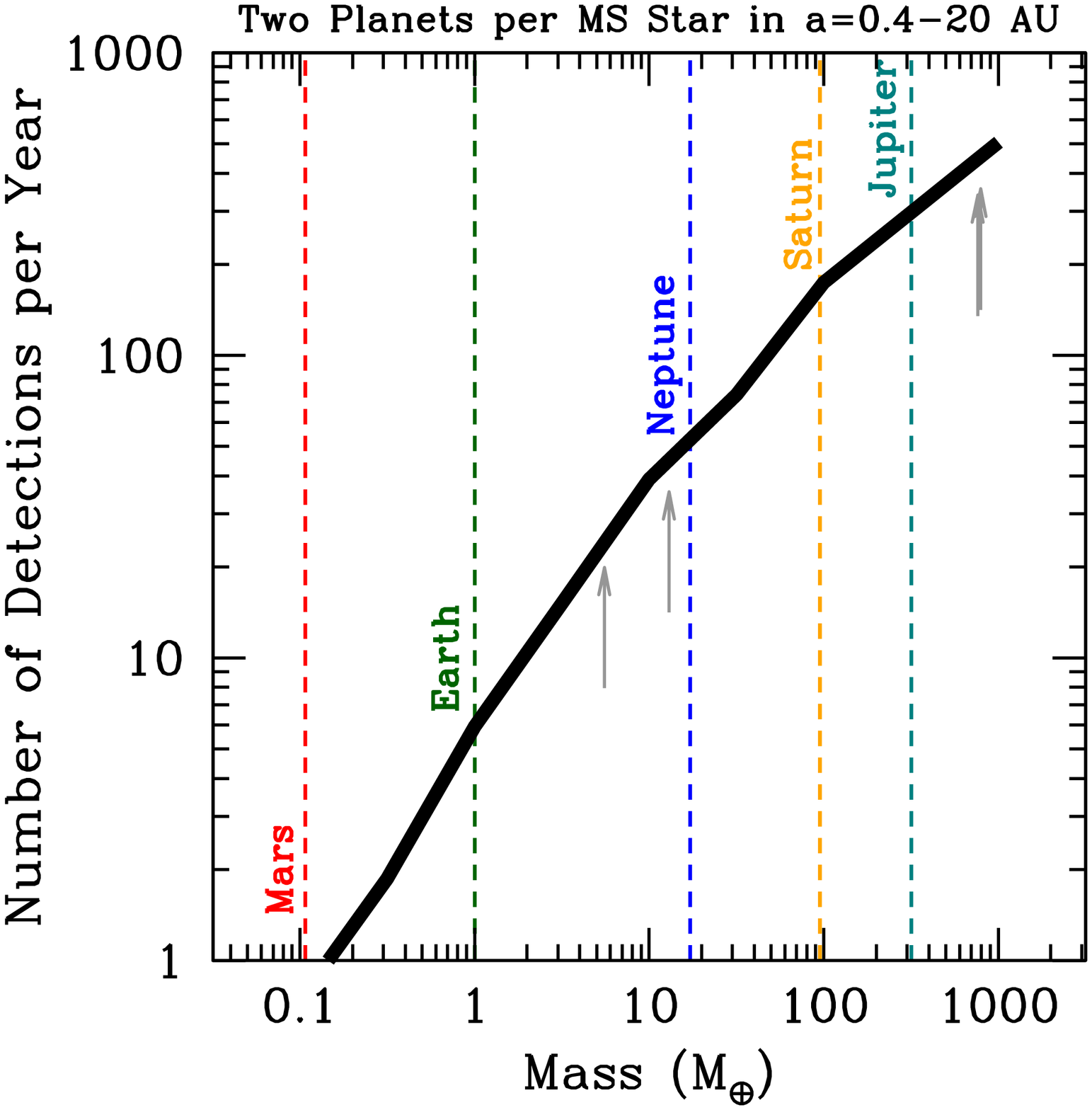}
\caption{
Expectations from a next-generation ground-based microlensing survey.
These results represent the average of two independent simulations
which include very different input assumptions 
but differ in their predictions by only $\sim 0.3$ dex. 
{\bf Left} Number of planets detected per year 
assuming every main-sequence (MS) star has a planet
of a given mass and semi-major axis.  
{\bf Right} Same as left panel, but assuming every MS has two planets distributed
uniformly in $\log(a)$ between 0.4-20~AU. The arrows
 indicate the masses of the four microlensing exoplanet detections.
}\end{figure}

In fact, a next-generation microlensing survey appears to be forming
spontaneously.  The MOA telescope already represents one leg of 
such a survey.  The OGLE team has been funded to
upgrade to a 1.7 deg$^2$ camera, which would represent the second
leg.  Astronomers from Korea, Germany, and
China are considering initiatives to secure
funding to build 1-2m class telescopes with wide FOV cameras in southern Africa or
Antarctica.  If these plans are realized, a next-generation survey
would effectively be in place that could realize a significant fraction of the
yields shown in Figure 2. 

Ultimately, however, the true potential of microlensing cannot be
realized from the ground.  Weather, seeing, crowded fields, and
systematic errors all conspire to make the detection of
planets with mass less than Earth effectively impossible from the
ground.  A space-based microlensing survey offers several advantages:
the main-sequence bulge sources needed to detect sub-Earth
mass planets are resolved from space, the events can be monitored
continuously, and it is possible to observe the moderately reddened
source stars in the near infrared to improve the photon collection
rate.  Furthermore, the high spatial resolution afforded by space
allows unambiguous identification of light from the primary
(lens) stars and so measurements of the primary and planet masses.

The expectations from a Discovery-class space-based microlensing
survey are impressive.  Such a survey would be sensitive to all planets
with mass $\ga 0.1M_\oplus$ and separations $a \ga 0.5~{\rm AU}$,
including free-floating planets. This range includes analogs to all
the solar system planets except Mercury. If every main-sequence star
has an Earth-mass planet in the range $1-2.5~{\rm AU}$, the survey
would detect $\sim 500$ such planets within its mission lifetime.
When combined with complementary surveys such as {\it Kepler}, such a
survey would yield an accurate and complete census of both bound and
free-floating planets with masses greater than that of Mars orbiting
stars with masses less than that of the Sun.  Such a census would
likely provide the ultimate test of planet formation theories.

\acknowledgements 
I would like to acknowledge the many people who have contributed to
the work described in this short review, and in particular Andy Gould,
David Bennett, Subo Dong, and my fellow members of the $\mu$FUN
collaboration.  I would like to thank Fred Rasio, Alex Wolszczan, and
the other organizers for the putting together an amazing
conference.

\end{document}